\documentstyle[tighten,twocolumn,prl,aps,psfig]{revtex}
\begin{document}

\title{A first-principles study of MgB$_{2}$ (0001) surfaces}

\author{Zhenyu Li, Jinlong Yang \thanks{%
Corresponding author. E-mail: jlyang@ustc.edu.cn}, J.G. Hou,
Qingshi Zhu}

\address{
Open Laboratory of Bond Selective Chemistry and Structure Research Laboratory,
University of Science and Technology of China, Hefei, Anhui
230026, P.R. China}

\date{\today}

\maketitle

\begin{abstract}

We report self-consistent {\it ab initio} calculations of
structural and electronic properties for the B- and Mg-terminated
MgB$_{2}$ (0001) surfaces. We employ ultra-soft pseudopotentials
and plane wave basis sets within the generalized gradient
approximation. The surface relaxations are found to be small for
both B- and Mg-terminated surfaces. For the B-terminated surface,
both B ${\sigma}$ and ${\pi}$ surface bands appear, while only one
B ${\pi}$ surface band exists near the Fermi level for the
Mg-terminated surface. The superconductivity of the MgB$_2$
surfaces is discussed. The work function is predicted to be 5.95
and 4.25 eV for the B- and Mg-terminated surfaces respectively.
The simulated scanning tunneling microscopy images of the surfaces
are not sensitive to the sign and value of the bias voltages, but
depend strongly on the tip-sample distance. An image reversal is
predicted for the Mg-terminated surface.

\end{abstract}


The surprising discovery \cite{nagamatsu} of superconductivity at
39K in MgB$_{2}$ has attracted great scientific interest. The
isotope effect experiments\cite{budko,hinks} indicate that
MgB$_{2}$ may be a BCS phonon-mediated superconductor, with Tc
above the commonly accepted limits for phonon-assisted
superconductivity. Band structure calculations \cite{an,kortus}
show that the Mg donates substantially its 3s electrons to the B
layer, and that the Fermi surface is derived mainly from B
orbitals with two-dimensional character. The superconductivity is
believed to be due to the strong coupling of the holes of the B
$\sigma$ bands to the in-plane B phonon lattice vibrations
(E$_{2g}$ modes) \cite{an}.

Being a simple, low-cost and high-performance binary intermetallic
compound, MgB$_{2}$ is a very promising candidate for
superconducting device as well as large-scale applications.
Experimentally, scanning tunneling spectroscopy measurements of
the surface of superconducting MgB$_2$ have been
reported\cite{karapetrov,rubio}, and the high-quality
c-axis-oriented epitaxial MgB$_{2}$ thin films have been grown
successfully on Al$_2$O$_3$ substrates by using a pulsed laser
deposition technique \cite{kang}.   However, there has not been a
theoretical study for the MgB$_{2}$ surfaces so far. In this
Letter, we report a first-principles study of MgB$_{2}$ (0001)
surfaces. Both B- and Mg-terminated surfaces are investigated. We
show that while the relaxations of the surface layers are small
for both surfaces, their electronic structures are quite
different.

Our calculations were carried out within the generalized gradient
approximation (GGA) \cite{pw91} using ultrasoft pseudopotentials and plane
wave basis sets \cite{vasp1,vasp2}. A \(1\times1\times15\) supercell geometry was
used to model the surfaces in which there are 15 atomic layers and
15 layers of vacuum (hereafter, 15-layer slab model) . The plane
wave cutoff is 257.2 eV for structural optimizations and 321.5 eV
for static electronic structure calculations. Brillouin zone
integrations were performed on a grid of \( 13\times 13\times 1 \)
Monkhorst-Pack \cite{mp} special points. During the structural
optimizations, we fixed the central 3 atomic layers in the bulk
configuration and allowed all other atoms in the supercell to move
until all forces vanish within 0.02 eV/\AA. We have checked some
our calculations with a 17-layer slab model, and found the
15-layer slab model is sufficient for our studies.

The bulk MgB$_{2}$ consists of alternating hexagonal layers of Mg
atoms and boron honeycomb layers. Our bulk calculations give the
lattice constants of \( a=3.067\AA \), \( c=3.515\AA \), which are
in good agreement with experimental and other theoretical results
\cite{kortus,lipp}. We used this bulk geometry as the initial
geometry for the surface optimizations. Table 1 lists the results
for the surface relaxations, given in terms of the change
$\Delta_{ij}$ of the interlayer distance in percent of the
distance in the bulk. From this table, one can see the relaxations
for both the B- and Mg-terminated surfaces are rather small (less
than 4\%). The relaxation affects up to five atomic layers for the
B-terminated surface, while it mainly localizes on the first two
layers for the Mg-terminated surface. For the B- and Mg-terminated
surfaces, an inward relaxation of the top layer by -2.1\% and
-3.7\% and an outward relaxation of the second layer by 2.0\% and
1.2\% are observed. The energy gains due to the relaxation are
2.31 and 2.38 eV for the B- and Mg-terminated surfaces
respectively.

The surface electronic structure of the B- and Mg-terminated
MgB$_{2}$ (0001) surfaces at the optimized geometries is shown in
Fig.1 together with the projected band structure (PBS) of the bulk
MgB$_{2}$.  From Fig. 1, we can see that for the B-terminated
surface, there is one surface band along each of the bulk bands.
The dispersion of these surface bands is very similar to that of
the bulk bands, and they can be sorted into the bonding ${\sigma}$
(sp$_x$p$_y$) and nonbonding ${\pi}$ (p$_z$) bands of the surface
B atoms accordingly. For the Mg-terminated surface, however, there
is only one surface band outside the PBS. This band mainly results
from the p$_z$ orbitals of B atoms in the second surface layer.

The above results can be understood by plotting the charge density
distribution. The charge density contour maps for the x-z plane
and x-y plane at B layers are shown in Fig. 2. In good agreement
with other calculations for the bulk, we found that electrons are
transferred form Mg orbitals to B orbitals, but they are not well
localized at the B sites, instead they are distributed over the
whole crystal. The MgB$_2$ is essentially metallic boron held
together by covalent B-B and ionic B-Mg bonding. At the
B-terminated surface, both in-plane (sp$_x$p$_y$) and out-plane
(p$_z$) electrons of the surface B layer are slightly
redistributed due to lack of one nearest neighbor (NN) Mg layer.
This kind of electronic redistribution results in the appearance
of the  B ${\sigma}$ and ${\pi}$ surface bands. At the
Mg-terminated surface, however, the second surface B layer has two
NN Mg layers as in bulk, and only its out-plane (p$_z$) electrons
are perturbed by the electron redistribution of the surface Mg
layer, leading to a B ${\pi}$ surface band.

Figure 3 shows the total densities of states (DOS) and  layer
resolved projected DOS for the B- and Mg-terminated MgB$_2$
surfaces together with those for the bulk.  While the total DOS
curves are quite similar overall, there are differences in the
layer resolved project DOS. The surface-specific features in DOS
disappear slowly for the B-terminated surface, while they vanish
rapidly for the Mg-terminated surface. For the Mg-terminated
surface, those features are mainly localized in the (Mg, B)
surface layers, the DOS of the first (Mg, B) sub-surface layers
are already virtually identical with those of the bulk. Even in
the second (B, Mg) sub-surface layers for the B-terminated
surface, however, we still can see those features. The total DOS
at the Fermi level ($N(0)$) is calculated to be 0.42 and 0.37
states/(eV$\times$ B atom) for the B- and  Mg-terminated surfaces
respectively. Both are larger than the $N(0)$ of the bulk (0.36
states/(eV$\times$ B atom)).

To discuss qualitatively the superconductivity of the MgB$_2$
surfaces, we calculated the layer resolved zone-center E$_{2g}$
(in-plane displacements of borons) phonon frequencies. They are
596 cm$^{-1}$ and 479 cm$^{-1}$ for the first surface B layer of
the B- and Mg-terminated surfaces respectively. These values are
bigger than that of the bulk-like B layer (~476 cm$^{-1}$). From
our $N(0)$ and  E$_{2g}$ results, we expect that the
superconductivity of the MgB$_2$ surfaces is somewhat
strengthened. This conclusion is contrary to the conjecture of a
weakened surface layer \cite{schmidt}. The conjecture was resulted
from that the measured superconducting energy gap is less than the
weak-coupling BCS value of 5.9 meV for the bulk T$_c$ of 39
K\cite{karapetrov,rubio,schmidt}. However, we notice that the
measurement might be done on chemically modified surfaces, not on
the ideal surfaces.

We also calculated the work function of the MgB$_{2}$ surfaces in
terms of the difference between the self consistent potential in
the vacuum and the fermi level. As listed in Table 1, the B-
termnated surface gives a work function of 5.95 eV, while the Mg-
termnated surface is 4.25 eV. This result indicates that the
electrons on the B-terminated surface are more stable than Mg-
terminated surface.

The scanning tunneling microscopy (STM) is a widely used tool in
surface research. Here we simulated the STM images for the
MgB$_{2}$ surfaces using the Tersoff and Hamann theory \cite{tersoff}. By
assuming an asymptotically spherical tip and taking the limit of
small applied bias voltage, the tunneling current was reduced to
the local density of states (LDOS) of the surface at the point
probed by the tip:

$$I(x,y)\propto \rho (x,y)=\sum _{i}|\Phi _{i}(x,y)|^{2}\delta
(E_{i}-E_{F})~~~~~~~~ (1)$$

\noindent We found while the simulated images are not sensitive to
the sign and value of the bais voltages, they depend on strongly
the distance  between the tip and sample. Fig. 4 shows some
simulated images with different tip-sample distances at +0.5 V.
From these figures, one can see that at the Mg-terminated surface,
the STM images show a ''bright-dark" transition. At a short
distance, the bright area of the image is corresponding to Mg
sites in the surface layer. At a long distance, the bright area is
changed to the B sites of the second surface layer. This reversal
contrast of the STM images at different tip heights can be
understood by Eq.(1). In Eq.(1), the tunneling current is
determined by the space and energy windows simultaneously. At the
Mg-terminated surface, the energy contributions to the tunneling
current mainly come from the B layer while the surface Mg layer
gives the most space contributions when the tip is close to the
surface.

In summary, we have studied the structural and electronic
properties of the B- and Mg-terminated MgB$_{2}$ (0001) surfaces.
The surface relaxations are found to be small. The surface bands
are all along the bulk bands. For B-terminated surface, both
p$_{\sigma}$ and p$_{\pi}$ surface bands are exist, but for Mg-
terminated there is only a p$_{\pi}$ surface band. Compared with
the bulk, both the two surfaces are found to increase the total
DOS at the Fermi level. The superconductivity of the MgB$_2$
surfaces is found to be somewhat strengthened. The work function
is predicted to be 5.95 and 4.25 eV for the B- and Mg-terminated
surfaces respectively. The simulated STM images are found to be
sensitive to the tip-sample distances. For the Mg-terminated
surface we even observed an image reversal for different
tip-sample distances.

We thank Dr. D.M. Chen of the Rowland Institute for Science for
many helpful discussions. This work was partially supported by the
National Project for the Development of Key Fundamental Sciences
in China (G1999075305), by the National Natural Science Foundation
of China, by the Foundation of Ministry of Education of China, and
by the Foundation of the Chinese Academy of Science. The HPCC,
NSC, and SC\&CG Laboratory of USTC are acknowledged for
computational facilities.

\begin{figure}
\caption{Surface band structure of (a) the B-terminated surface
 and (b) the Mg-terminated surface. The projected band structure of the bulk MgB$_{2}$
is shown by shadow areas. }
\label{fig1}
\end{figure}

\begin{figure}
\caption{Charge density contour maps: (a)the x-z plane map of the
B-terminated surface, (b) the x-z plane map of the Mg-terminated
surface, and (c) the x-y plane maps at the first surface B layers
of the B- and Mg-terminated surfaces and at the B layer of the
bulk respectively. The contours are in the unit of electrons/$\AA
^3$.} \label{fig2}
\end{figure}

\begin{figure}
\caption{ Densities of states (DOS) for the surfaces and bulk.
(a1)-(a3) are the total DOS of the B-terminated surface,
Mg-terminated surface, and bulk. In (a3), we also plot the
projected DOS for the B and Mg atomic layers in the bulk.
(b1)-(b3) are the layer resolved projected DOS of the first,
second and third (B, Mg) layers from the top surface of the
B-terminated surface respectively. (b4)-(b6) are the counterparts
for the Mg-terminated surface. The bulk DOS was calculated on a
\(19\times19\times19\) Monkhorst-Pack k-point grid, while the DOS
of slab models was obtained on a \(19\times19\times2\) k-point
grid. } \label{fig3}
\end{figure}

\begin{figure}
\caption{Simulated STM images of MgB$_{2}$ (0001) surfaces at
+0.5eV: (a) B-terminated surface with the tip-sample distance
($d$) of 5 \AA, (b) B-terminated surface with $d$=3\AA, (c)
Mg-terminated surface with $d$=5\AA, and (d) Mg-terminated surface
with $d$=3\AA.}
\label{fig4}
\end{figure}

\begin{table}
\caption{Relaxation of MgB$_{2}$ (0001) surfaces (change $\Delta
d_{i,j}$ of the interlayer distances in percent of the distance in
the bulk) and work function ($\Phi$) in eV. }
\label{table1}
\begin{tabular}{ccccccc}
&
 $\Delta_{12}$ &
 $\Delta_{23}$ &
 $\Delta_{34}$ &
 $\Delta_{45}$ &
 $\Delta_{56}$ &
 $\Phi$\\
\hline
B-terminated&
 -2.1\%&
 2.0\%&
 0.9\%&
 -1.8\%&
 0&
5.95\\
Mg-terminated&
 -3.7\%&
 1.2\%&
 0.2\%&
 0.5\%&
 -0.3\% &
4.25\\
\end{tabular}
\end{table}

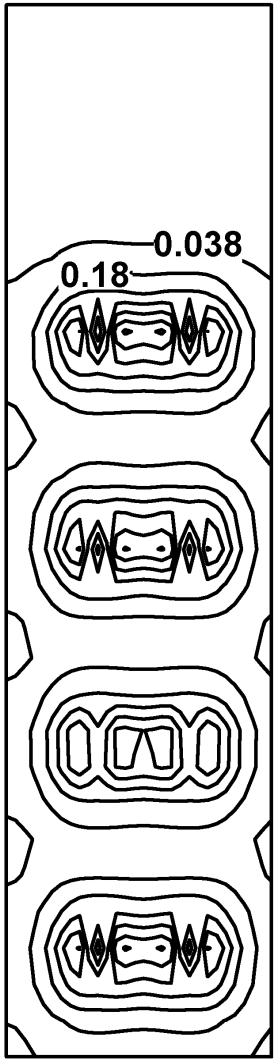
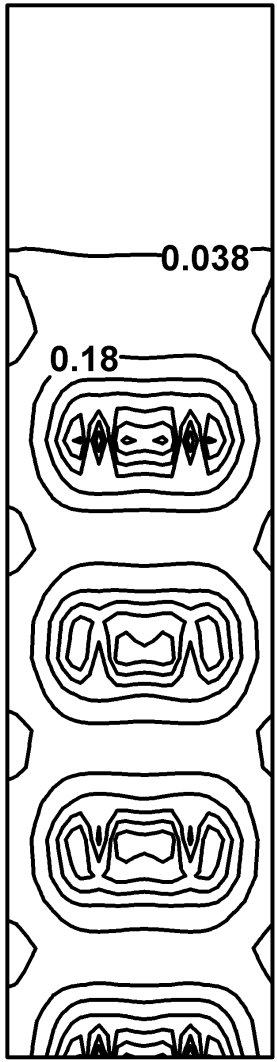
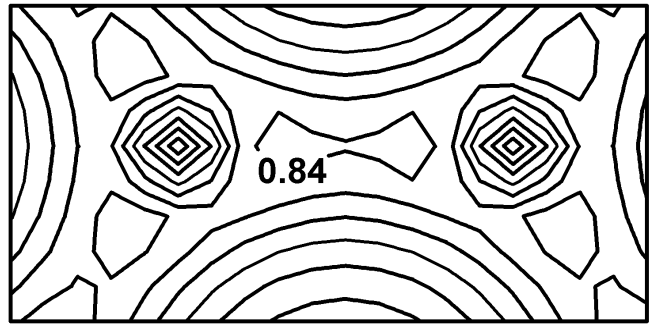
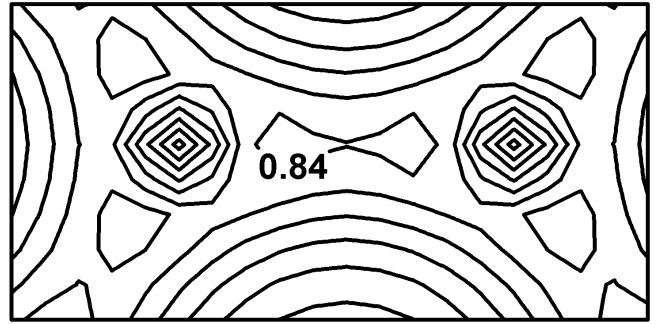
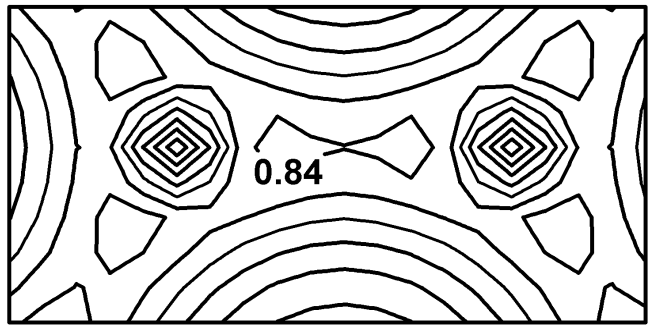
\end{document}